# In situ atomic force microscopy depth-corrected 3-dimensional focused ion beam based time-of-flight secondary ion mass spectroscopy: spatial resolution, surface roughness, oxidation

Lex Pillatsch[a,b], Szilvia Kalácska[a*], Xavier Maeder[a], Johann Michler[a]

[a]Empa, Swiss Federal Laboratories for Materials Science and Technology, Laboratory of Mechanics of Materials and Nanostructures, CH-3602 Thun, Feuerwerkerstrasse 39. Switzerland
[b]TOFWERK AG., CH-3600 Thun, Uttigenstrasse 22 Switzerland

*Corresponding Author's *E*mail address: szilvia.kalacska@empa.ch, Tel.: +41 58 765 626, Fax: +41 58 765 6990

**Abstract**

Atomic force microscopy (AFM) is a well-known tool for studying surface roughness and to collect depth information about features on the top atomic layer of samples. By combining secondary ion mass spectroscopy (SIMS) with focused ion beam (FIB) milling in a scanning electron microscope (SEM), chemical information of sputtered structures can be visualized and located with high lateral and depth resolution. In this paper, a high vacuum (HV) compatible AFM has been installed in a TESCAN FIB-SEM instrument that was equipped with a time-of-flight secondary ion mass spectroscopy (ToF-SIMS) detector. To investigate the crater's depth caused by the ToF-SIMS sputtering, subsequent AFM measurements were performed on a multilayer vertical cavity surface emitting laser (VCSEL) sample. Surface roughness and milling depth were used to aid accurate 3D reconstruction of the sputtered volume's chemical composition. Achievable resolution, surface roughness during sputtering and surface oxidation issues are analysed. Thus, the integration of complementary detectors opens up the ability to determine the sample properties as well as to understand the influence of the analysis method on the sample surface during the analysis.

1. **Introduction**

Focused ion beam secondary ion mass spectroscopy (FIB-SIMS) is a well-known surface analysis technique [ Dunn 1999][ Giannuzzi 2005][ Giannuzzi 2011]. The measurement principle is based on locally sputtering the sample by impacting ions while collecting and selecting the sputtered ions according to their mass. The sensitivity of SIMS makes it suitable for locating trace elements important to detect impurities in semi-conductors and metals. SIMS is also capable of analysing the chemical composition of multilayer structures [ Whitby 2012] with an excellent depth resolution. The remarkable depth resolution comes from the localization of sputtered particles that are limited to the few top surface layers [ Pillatsch 2018][ Priebe 2019]. Although SIMS is extremely surface sensitive, it is not possible to determine the depth of the sputtered particle as the sputter speed depends on the chemical and crystallographic composition of the sample and the projectile. Therefore, the elemental composition of the sputtered volume is usually represented as signal per sputtered frame. This often leads to a false impression of the depth-represented elemental composition. Depth profiling of the crater after SIMS measurement gives only information about the average sputtering speed of the analysed volume, which can vary significantly because of the different sputtering speed of the analysed structures at various crater depths.

An additional analysis method is required to calibrate the depth of the sputtered crater and to analyse the roughness of the crater bottom. This information is provided by atomic force microscopy (AFM). When combining AFM with SIMS, the AFM needs to be vacuum compatible as SIMS measurements are done in high vacuum (HV). Breaking the vacuum for AFM measurement needs to be avoided when alternating SIMS and AFM measurements as surface oxidation can occur. A suitable HV AFM setup that is easy to implement is based on a tuning fork AFM where the tip vibration is controlled and recorded by tracking the variations in the electrical impedance of the tuning fork [ Akiyama 2006][ Akiyama 2003a][ Akiyama 2010][ Akiyama 2003b]. No optical setup is required for this type of AFM which makes it easy to adjust in vacuum.

An advantage of *in situ* AFM consists in a higher quality factor (Q-factor) compared to in-air AFM. The damping force from ambient air acting on the vibration of the tip is missing in vacuum which allows a free vibration of the tip, hence the Q-factor is increased. The consequence of a high Q factor is faster response reaction to variation of the tip-to-surface interaction [ Kushmeric 2016] and thus the sensitivity to surface variations is higher.

At ambient pressure, the AFM can be driven either in amplitude mode (AM-AFM) by keeping the amplitude of the tip vibration constant, or in frequency mode (FM-AFM) by keeping the frequency of the tip vibration constant. Both methods are used in tapping mode where although the tip is close to the surface, but any actual contact between surface and tip should be avoided. For AM-AFM, the frequency of the tip changes according to the interaction force between the tip and the surface. The frequency shift of the tip is proportional to the scan speed and inversely proportional to the quality factor (or Q-factor) [ Kushmeric 2016][ Rodríguez 2003]. Apart from measuring the surface height, AM-AFM mode also provides information about the surface composition by tracking the phase shift between the vibration frequency and the excitation frequency of the tip which is proportional to the tip-surface interaction force [ Bayat 2008]. A drawback of this measurement method is the link of the Q-factor to the scan speed. In HV conditions, the elevated Q-factor leads to a reduced scan speed in order to keep the tip vibration in an acceptable frequency range for a given tip vibration amplitude. This will slow down the scan speed to an unacceptable level.

For FM-AFM, the Q-factor is independent from the scan speed. The frequency detection is not affected by the amplitude of the tip vibration. This allows to maintain high scan speeds while recording precise information of the surface height. As the phase shift cannot be measured, the composition of the surface cannot be tracked by this method. Nevertheless, the FM-AFM precision in terms of depth measurement is higher than with AM-AFM.

The visualization of the crater is done by SEM after the last AFM scan. The SEM image is useful to determine and overlay the scan area of the earlier AFM and SIMS measurements and to confirm the measured AFM data after surface sputtering.

In this article, we present a scanning strategy to alternate AFM and SIMS measurements in order to get the depth information about the chemical composition of a multilayer sample. Advantages and drawbacks of the acquisition are pointed out and limits of the measurement are discussed.

## 2. Materials and Methods

A FIB-SEM instrument developed by TESCAN served as platform for the SIMS-AFM integration. The sample is sputtered by a gallium (Ga) FIB. The surface damage induced by the impacting Ga ions is visualized by SEM pictures. The elemental composition of the sample is measured by an orthogonal ToF-SIMS instrument [ Alberts 2014] developed by TOFWERK AG. With the HV compatible AFM the surface roughness and the sputter depth in between the SIMS scans are measured.

### 2.1. AFM setup

Figure 1. shows the AFM with a tuning fork tip that was designed to fit in the FIB-SEM instrument. For the so-called "Akiyama tips" the silicon (Si) tip apex is attached between two prongs [ Akiyama 2010]. The deflection of the tip, orthogonal to the vibration plane of the prongs, is induced by periodical changing of the mechanical stress in the legs of the cantilever. The tuning fork itself acts as an oscillatory force sensor and dominates the frequency and the amplitude of the tip vibration. The interaction of the tip with the surface is measured by the piezoelectric current of the tuning fork. The resonance frequency of the tip is about 45 kHz with Q-factor measured in air of about 1000. In vacuum, the tip vibration is not damped by air. The Q-factor is 4-5 times higher which has a favourable impact on the reaction precision of the tip height. The force constant of the cantilever is 5 N/m. The apex radius of the tip is about 15 nm. AFM measurements are performed in FM-AFM tapping mode. The compactness of the AFM setup allows for mounting the AFM tip between the extraction optics of the ToF-SIMS, the FIB and the sample surface. The sample is mounted on the AFM piezo scanner with a maximum scan range of 10 μm × 10 μm. For a 512 px × 512 px scan, so a lateral resolution of 20 nm and a scan speed of 1.5 s/line, an AFM scan takes about 13 minutes. The height adjustment of the tip is done by a piezo scanner below the tip with a maximum movement range of 5 μm. The coarse motion of the tip is done by a tripod setup of slip stick actuators. The coarse motion is independent of the AFM scanner and serves only to set the tip at the area of interest and so to intersect the tip scan area with the scan area of the SIMS. During the actual AFM scan, the tip height is regulated by a piezo actuator bellow the AFM tip.

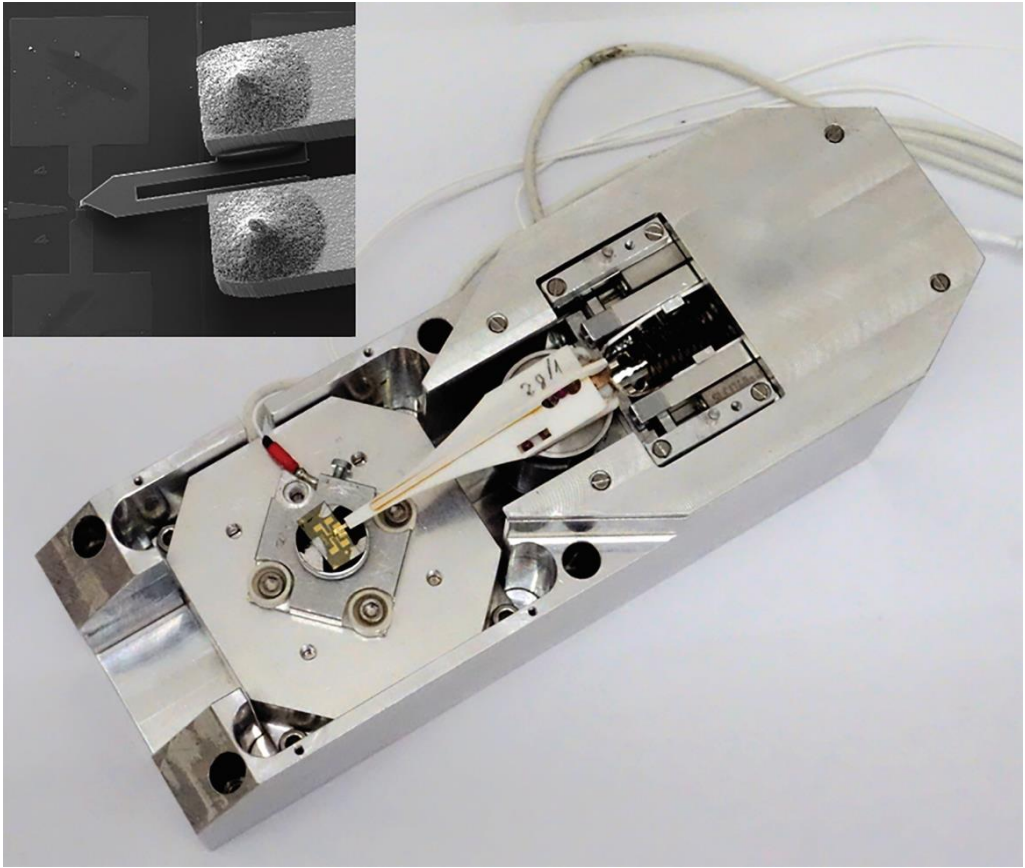

Figure 1: HV AFM setup with the sample installed on the 10 μm × 10 μm range scanning table. The adjustment of the tip height is done by a piezo manipulator below the tip. Coarse motions to position the tip at the area of interest are realized by slip stick nanomanipulators. The inset picture shows a secondary electron image of the tip close to the surface.

## 2.2. SIMS

The SIMS measurements are done with a high-resolution ToF-SIMS detector from TOF-WERK AG, Thun, Switzerland [Whitby 2012]. The mass resolution of the orthogonal ToF-SIMS is $\Delta m/m > 3900$ Th/Th. The sample is sputtered by the Ga+ beam perpendicular to the sample surface. Beam energy of 20 keV is chosen for the experiment in order to guarantee a lateral resolution of about 50 nm. The extraction voltage for the secondary ions applied to the extraction nozzle of the ToF-SIMS is limited to < 200 V, while the sample itself is grounded. The distance between the extraction optics and the sample surface is 9 mm. Because of the weak electric field, the AFM tip can stay at a distance of about 20 μm from the scanning area without disturbing the ion extraction of the TOF-SIMS. Therefore, it is possible to make fast alternation between SIMS and AFM scans with a repetition precision in the range of 100 nm that is limited by the coarse motion for the tip positioning.

## 2.3. SEM

The exact location of the tip and the distance between the sample surface and the AFM tip is visualized by SEM. As the magnification of the SEM is higher by orders of magnitude than for optical microscopes, approaching and positioning of the tip at the area of interest is easy. Furthermore, damage of the apex of the tip can be noticed prior to AFM measurements. If the AFM parameters are adjusted adequately, the tip apex will last during several AFM scans which is required for a correct AFM-SIMS reconstruction of the sample. The scan area of the SIMS measurement is referred to the distance of the crater done by the FIB beam. The initial position of the SIMS scan area and the AFM scan area can therefore be determined. The SEM is also applied to visualize the sputter location after the SIMS-AFM measurement to confirm the correctness of the AFM data.

*2.4.* **Sample**

The concept of depth calibration of the SIMS depth profile with direct depth measurements by AFM is demonstrated on a multilayer vertical cavity surface emitting laser (VCSEL) [Lyytikäinen 2009]. The exact layer structure of the sample is shown in Figure 2 a). This sample was chosen because of its initial surface flatness and its elemental composition being favourable for SIMS measurements.

The well-defined layer structure with the discrete border limitation is ideal to measure the intermixing due to collision cascades induced by impacting ions. Figure 2 b) shows the SIMS depth profile of $^{27}$Al, $^{69}$Ga and $^{115}$In as a function of the SIMS frames.

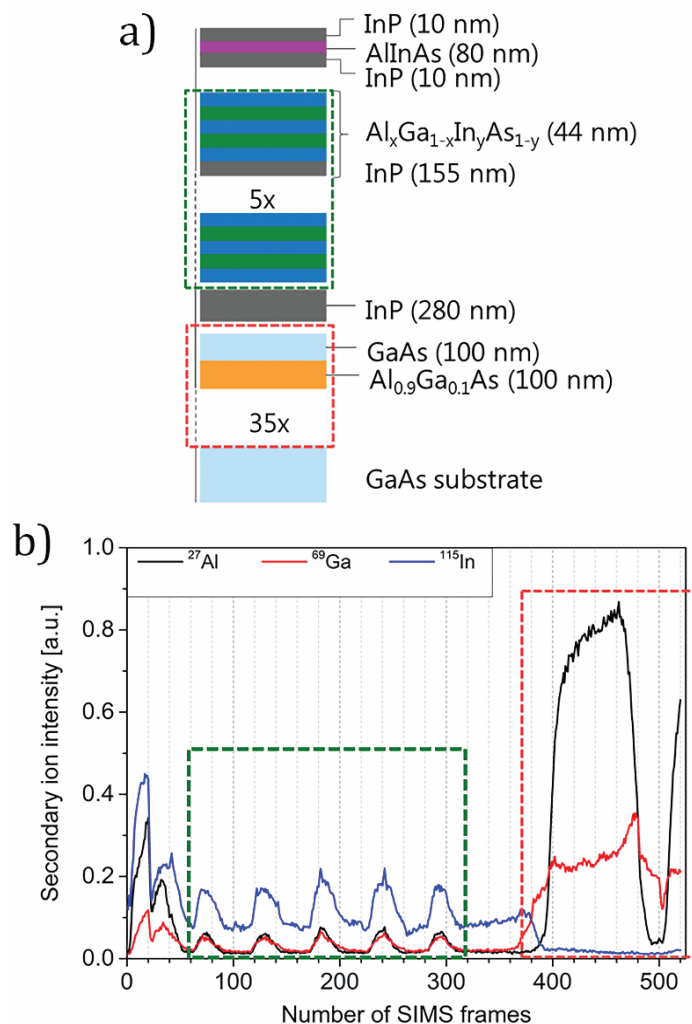

Figure 2: a) Layer structure of the VCSEL sample composed by 5 times a set of an aluminium rich layer structure separated by an indium phosphide (InP) layer, followed by 35 times alternating gallium arsenide (GaAs) and aluminium gallium arsenide (AlGaAs) layer structure. b) SIMS depth profile of the VCSEL sample showing the Al-rich structure and one of the AlGaAs layers.

### *2.5.* **Numerical control of the SIMS-AFM combination**

The SIMS-AFM combination is either used to determine the surface roughness and the crater depth at a specific SIMS frame or the SIMS-AFM scans are alternated, giving a 3D data set with the real sputter depth. The alternation of the SIMS-AFM scans is controlled through a Python script. The script only takes over control when switching between SIMS and AFM, so to start and stop a SIMS acquisition, to relocate the AFM tip before and after a SIMS acquisition (Figure 3 a) and to launch the AFM scans. The SIMS- and AFM-specific parameters such as scan range, scan speed, resonance frequency of the AFM tip are all set in the software of the corresponding analysis tool. New SIMS data acquisition is launched after each AFM acquisition set, resulting in several SIMS data sets that can be handled together by an external software. The periodicity of alternation between SIMS and AFM is determined by the number of SIMS frames per SIMS data set.

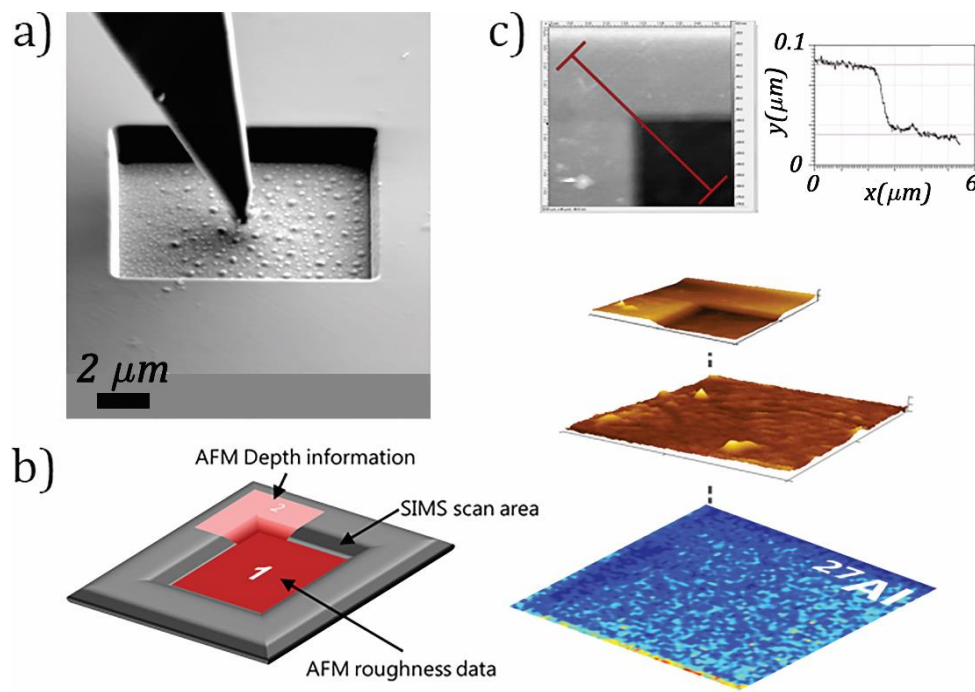

Figure 3: a) Centering of the tip in the SIMS crater to calibrate the zero level of the AFM tip. b) SIMS-AFM scanning strategy. After the SIMS data acquisition, the roughness of the sputtered crater is scanned by AFM (Area 1). In a second scan (Area 2), the crater edge is scanned in order to determine the crater depth. c) The information about the crater depth, crater roughness and the $^{27}$Al distribution measured by SIMS are represented.

### *2.6.* **SIMS-AFM scanning strategy**

Realizing the AFM measurements are the most time consuming. As the AFM tip is scanning line by line with a defined scan speed, the scan time increases with the power of 2 with the increase of the scan area for a given lateral resolution. For the analysed sample a SIMS scan area of 10 μm

× 10 µm is enough to get the required information about the chemical composition of the different layers. This scan size is on the edge of the scanning range of the AFM setup.

A time efficient scanning strategy leading to collect depth and roughness information at the centre of the crater consists of two AFM scans (Figure 3 b and c). The inner part of the SIMS crater is measured first by an AFM scan to record the roughness of the crater bottom. A second AFM scan with the crater edge in the middle of the scan area measures the sputter depth of the crater. By overlaying both AFM scans, all the required information can be extracted. Crater depths were determined by the Gwyddion software [Nečas 2012].

### 3. Results

#### 3.1. Surface oxidation

One AFM scan took about 25 min time. With a pressure p = $10^{-6}$ mbar in the analysis chamber and by assuming that the residual gas in the FIB-SIMS-SEM instrument consists only of $N_2$ molecules (the chamber is vented with nitrogen when changing the sample) the number of impinging atoms at the analysis surface during the AFM scan is calculated by [Jeans 2009]:

$$N_{mol} = \frac{p}{\sqrt{2\pi m k_B T}}, \qquad (1)$$

where $m$ is the molecular mass of $N_2$ molecules, $k_B$ is the Boltzmann constant an $T = 298$ K is the ambient temperature, resulting $N_{mol}$ = 2.9×$10^6$ molecules/µm²s impinging nitrogen molecules. Within 25 minutes, 4.3×$10^9$ molecules/µm² will imping a surface unit. This is sufficient to cover the surface. Whereas only $N_2$ molecules are assumed to be in the residual gas, in reality, a non-negligible fraction of chemically more reactive molecules ($O_2$, $H_2O$, $C_2$, ...) are contained as well in the residual gas. These reactive molecules will react with the surface during the AFM scan. Of course, only a given fraction of impinging molecules can react with the surface and therefore have an effect on the chemical surface state.

The ionisation probability of sputtered particles (and so the secondary ion signal) depends on the chemical state of the surface. The effect of surface oxidation enhances the secondary ion formation, and as a result, the signal of the first SIMS frame recorded after an AFM scan is significantly higher. Already at the subsequent SIMS scan, most of the oxidized surface is sputtered away, resulting in SIMS signal drop, appropriate to a clean surface. In Figure 4. the effect of surface oxidation on the SIMS signal after the AFM scan (every twentieth SIMS frame) is visible.

Being aware of this effect, the signal spikes are eliminated by averaging over the last secondary ion signal before and the ion signal of the second SIMS frame after an AFM scan in order to reflect the SIMS depth profile in absence of surface oxidation.

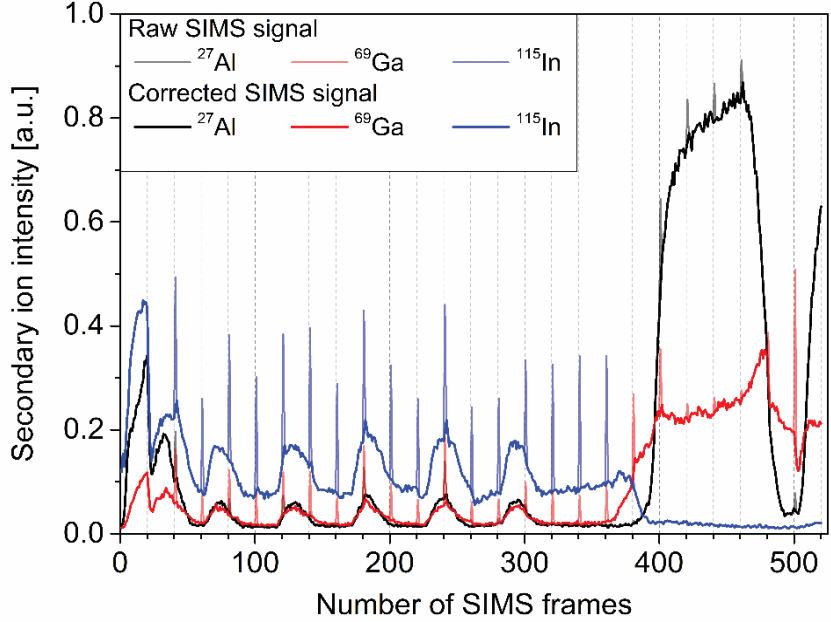

Figure 4: Effect of surface oxidation on the recombined depth profile of the $^{27}$Al, $^{69}$Ga and $^{115}$In SIMS signal. At every 20th SIMS frames, sputtering is interrupted for an AFM scan. The effect of surface oxidation during the AFM scan is visible by the enhanced secondary ion signal at the first SIMS frame after the AFM scan. Data correction is done by averaging the SIMS signal over the last SIMS scan before and the second SIMS scan after an AFM scan.

### 3.2. Sputter speed and roughness analysis by AFM

The depth profile of the VCSEL sample is recorded by SIMS. Beginning with an AFM scan, the SIMS depth profile is interrupted at every 20th frames to measure the sputter depth and the crater roughness by AFM.

Figure 5 shows the results of the combined SIMS-AFM measurement. The concentration variation of aluminium ions ($^{27}$Al), gallium ions ($^{69}$Ga) and indium ions ($^{115}$In) in the layer structure of the VCSEL is reflected by the SIMS depth profile. The sputter speed per 20 SIMS frames of the sample varies as a function of the local sample composition. It can be noticed that the sputter speed is highest on the falling slope of SIMS signals of the indium rich layers (AFM scan 3, 5 and 8) than on its corresponding rising slope (AFM scan 4, 6, 9, 12 and 15). This indicates that the sputter behaviour is influenced by the interlayer mixing. The sputtering of one of the Al-rich structure (AFM scan 7, 10 and 15) is lower than the sputtering speed at the AlGaInAs/InP interfaces, but still higher than the sputtering speed of InP (AFM scan 17 and 18).

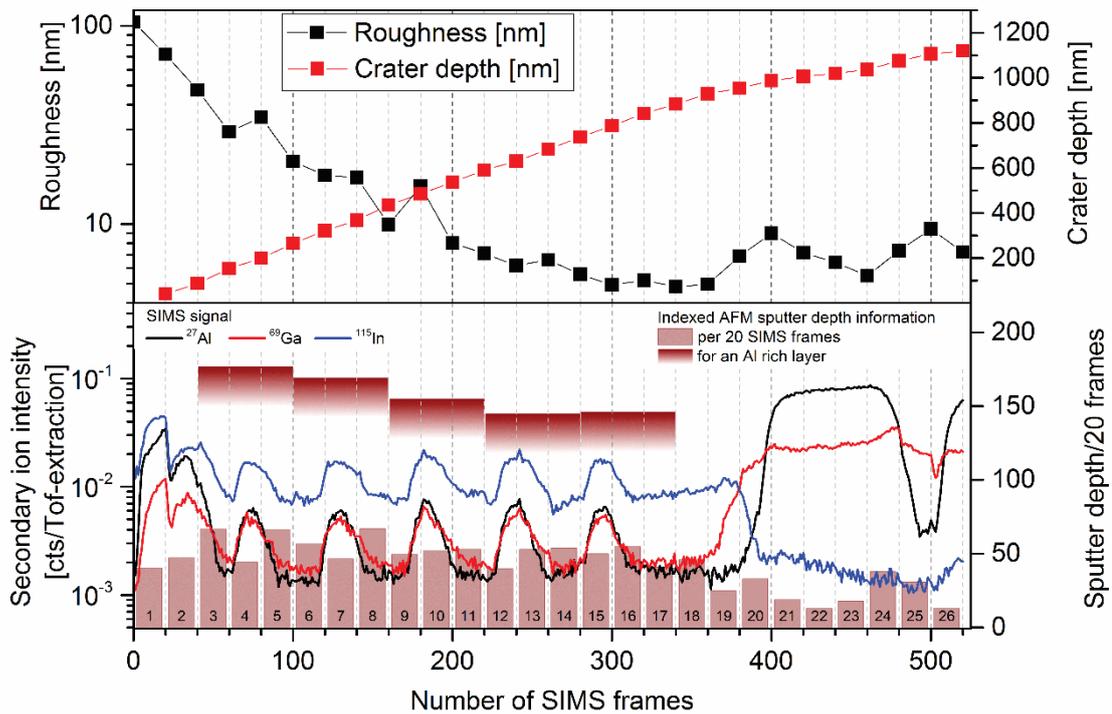

Figure 5: Top graph: Roughness and crater depth evolution as a function of the number of SIMS frames. Bottom graph: the $^{27}$Al, $^{69}$Ga and $^{115}$In depth profile is shown and related to the depth information measured by AFM. The sputter speeds of Al-rich layer structures are evaluated in terms of sputter depth.

As the sputter behaviour of the sample is unknown prior to measurement, it is not possible to synchronize the depth measurements with the SIMS signals. Nevertheless, it is possible to determine the consistency of the sputter speed of an Al-rich layer structure by summing the sputter depth of three consecutive AFM measurements incorporating an Al rich layer structure (Figure 5). For all 5 Al-rich structures the same amount of material is sputtered away. For the first Al rich layer structure, the sputter depth of 177 nm is influenced by the foregoing AlInAs/InP layers as well as by the initial surface roughness. The sputter depth of 170 nm of the second Al-rich structure is still influenced by the high roughness of the surface. With the decreasing roughness the sputter depth for an Al-rich layer levels off to about 145-155 nm for the 3–5 Al-rich structure. This corresponds to the distance between two Al-rich layers defined during the production of the sample.

The crater depth shown in the top graph of Figure 5 increases constantly up to the 18th AFM scan, just before reaching the AlGaAs-GaAs layer structure. The sputter speed for this layer structure defers noticeably from the previous layer structure, resulting in a lower sputtering behaviour, and thus in a lower sputter depth per 20 SIMS frames. In Figure 6. the correction of the sputter depth of the $^{27}$Al signal by AFM measurements is compared to the depth profile related directly to the sputter time of the crater.

When referring the $^{27}$Al signal to the sputter time, the layer thickness is reflected incorrectly due to the involved unknown sputter speed of the different layers. This is clearly visible in Figure 4 by the mismatch between the $^{27}$Al signal once represented as a function of the sputter time and once represented by a correct depth calibration of the sputtered crater.

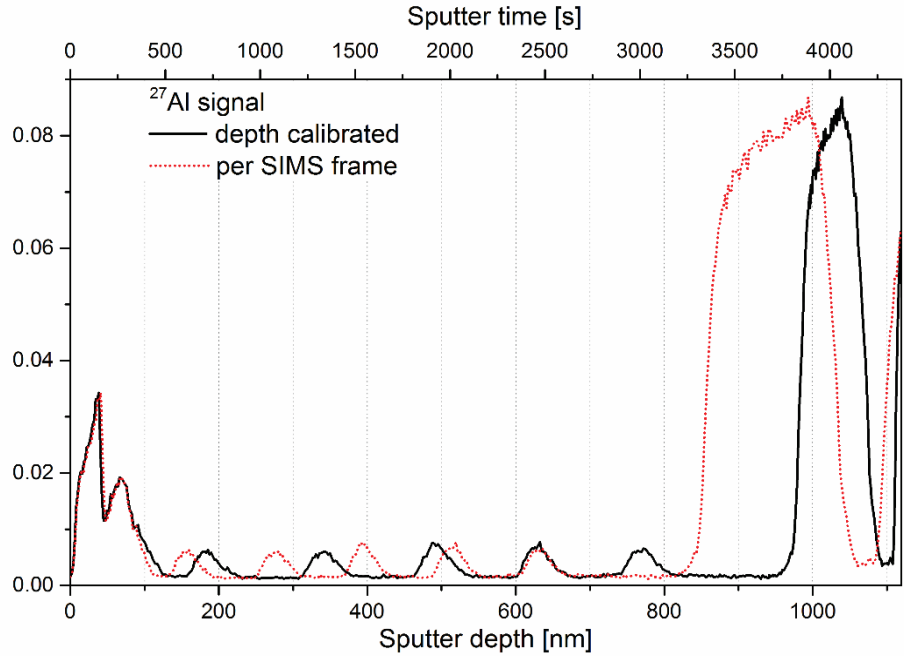

Figure 6: Overlay of the depth-corrected $^{27}$Al signal with the original $^{27}$Al signal plotted as a function of sputter time.

### 3.3. Surface roughness

Starting with a roughness of 100 nm of the pristine surface, the decreasing trend of the roughness is observed until the 18th AFM scan. The initial surface roughness is reduced by sputtering without adding new roughness by the impacting ions. The variation of the root mean square (RMS) roughness of the crater bottom for the AFM scans 18 to 26 is related to bubble formation at the surface (Figure 7).

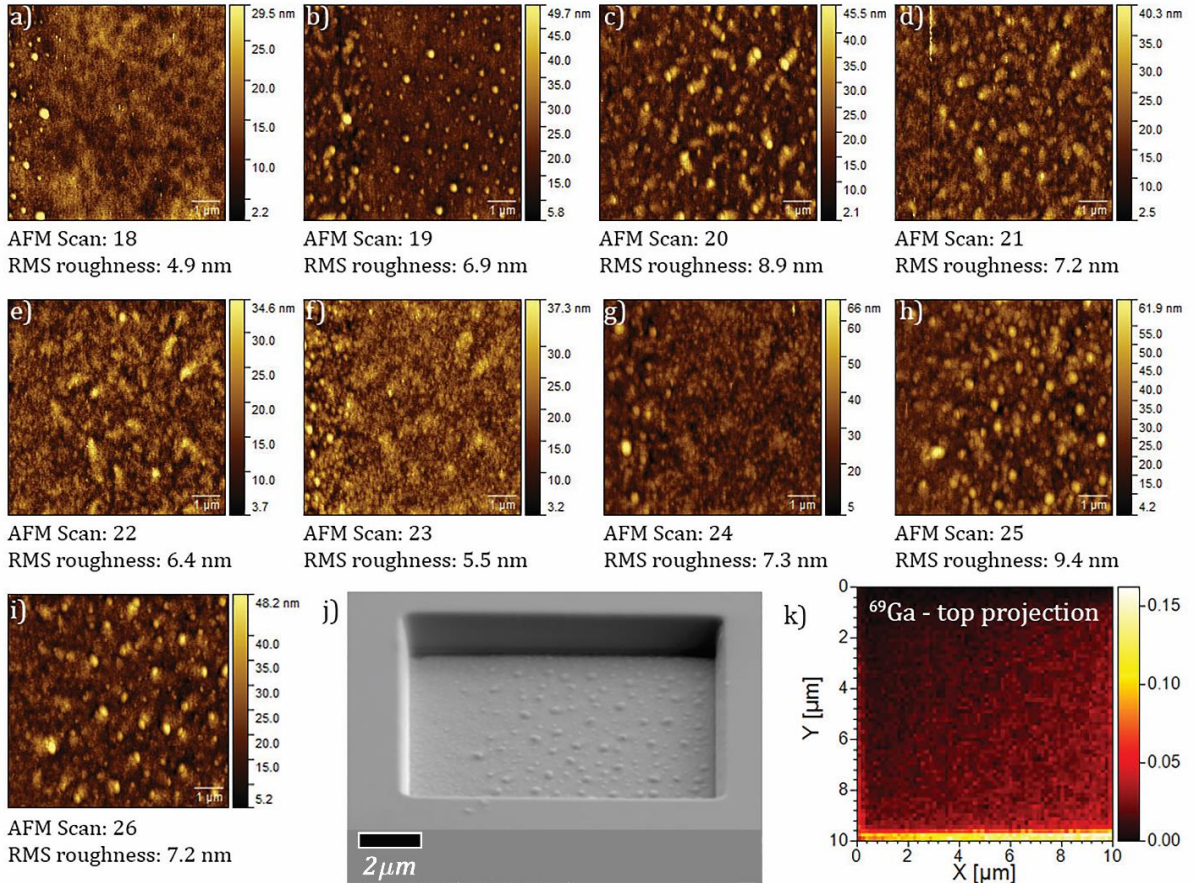

Figure 7: a-i) Surface roughness development from AFM scans 18 to 26. The RMS roughness values change due to bubble formation at the interfaces of the GaAs/AlGaAs layer. For the AFM pictures, the 8 μm× 8 μm middle part of the crater is shown. AFM height measurement at the crater edges are influenced by the dimensions of the AFM tip. The same bubble structure at the crater bottom is visible at the SEM picture taken after the last AFM scan (j), while no local increase in Ga could have been observed on SIMS images during droplet-formation (k).

On the left side of the 18th AFM scan, the formation of droplets starts already. When comparing the depth of the AFM scans with the SIMS data, droplet formation occurs mainly on the GaAs/AlGaAs interface. At the AFM scan 19 well separated droplets are formed. The low RMS roughness values are still dominated by the flat surface area in between the droplets. The RMS roughness increases at the AFM scan 20 due to the expansion of the droplets. When the GaAs/AlGaAs interface appears, the droplets are sputtered and a more uniform surface will appear. This leads to a reduction of the RMS roughness at the AFM scan 21-23, just before a new GaAs/AlGaAs interface appears and the droplet formation starts again to dominate the surface structure. The droplet formation continues, leading to an increase of the RMS values. As stated in the article of Wei et al. [Wei 2008] the mechanism of droplet formation on a GaAs surface can be attributed to preferential sputtering of As and clustering of the remaining excess of Ga, both form the Ga+ beam and from the substrate. They also identified the composition of the droplets by EDS and TEM to be pure Ga. According to TRIM2013 simulations the sputtering yield of As is 2.5 times higher than for Ga. On the other hand, no local increase in Ga could have been observed on SIMS images while droplets are formed. This can be explained by the movement of the droplets during sputtering and by Ga droplets clouded out by the Ga from the GaAs layer. The position of the droplets changes from frame to frame but the Ga signal of the ToF-SIMS image is averaged over 20 frames. This is enough time for the droplets to move over the entire sputtered area. Although the droplets are a local accumulation of Ga, the Ga concentration around the droplets is still in the 10$^{th}$ of percent range as Ga is present in the sample and the dose of Ga implanted by the Ga beam is high. The difference in Ga signal coming from droplets and from the area around the droplets might therefore only be minor.

## 4. Discussions

To compensate for the difference in sputtering rate for subsequent layers, the AFM data was used to create a depth-calibrated data set for the 3D reconstruction. Between each AFM measurement, the surface is sputtered during 20 frames to acquire ToF-SIMS data. The sputter speed of different layers is changing significantly but only sputtered depth between two AFM layers is measured, leading to an average sputter speed between two AFM measurements. For the 3D reconstruction, the ToF-SIMS data had to be split up again to 20nm thick slices. With a fixed 20 nm depth and a variating sputter speed, the number of ToF-SIMS frames is adapted to correspond to the 20 nm depth. With a 20 nm depth of each slice and a total sputter depth of the measurement of 1.1 µm, 56 individual slices containing an integrated and normalized signal intensity measured by ToF-SIMS are formed. The intensity distribution of the elements of the entire measurement with respect to the actual sputter depth is done by pilling up the 56 slices.

In Figure 8 the construction of three 3D model slices from the ToF-SIMS frames are shown. Voxel size in the 3D model is therefore 156 nm × 156 nm × 20 nm, as the ToF-SIMS mapping was binned to 64 pixel × 64 pixel resolution on a 10 µm × 10 µm area. In Figure 9 the 3D models reconstructed with Visage Imaging Amira 5.2.2 software are shown for the three studied ions.

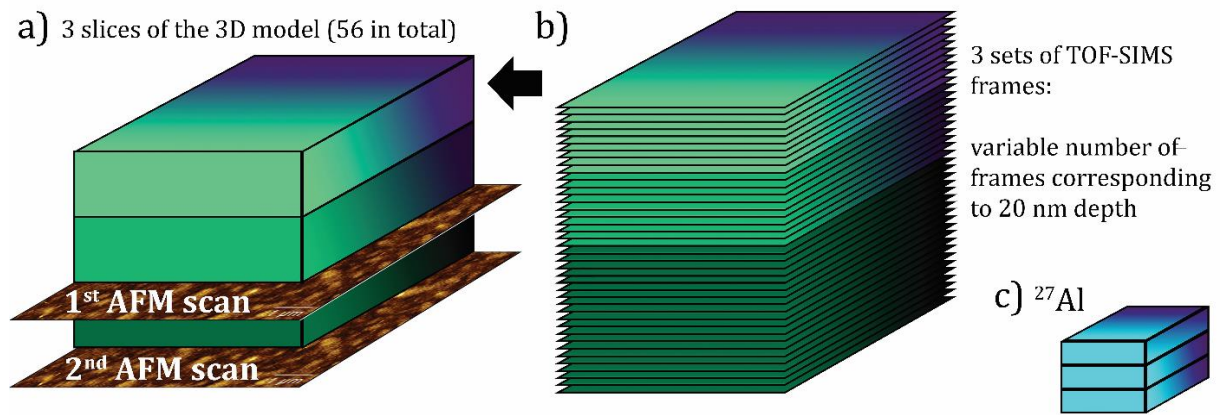

Figure 8: Sketch of the 3D model build-process based on variable number of ToF-SIMS frames. a) Slices used for the 3D reconstruction of the sample, b) Each slice consists of 20 nm thick layers, the amount of layers is adapted to the average sputter depth of one slice. c) An example showing that the 3D slices were constructed for the different ions, in this case $^{27}$Al.

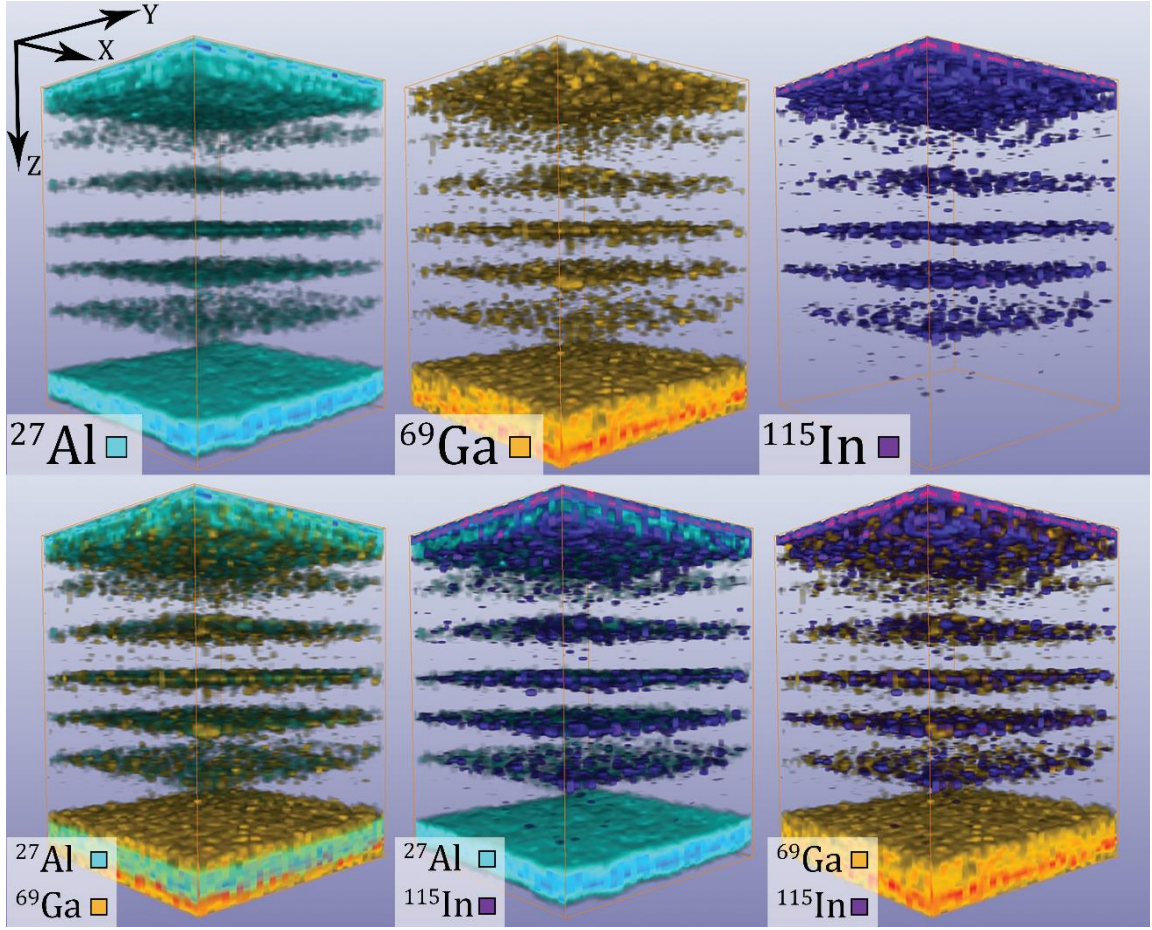

Figure 9: AFM Depth-corrected 3-dimensional distribution of $^{27}$Al, $^{69}$Ga and $^{115}$In ions. X and Y axes correspond to 5 μm lateral length, Z axis has been lengthened 6 times for better visualization, as in reality it only corresponds to 1.1 μm depth. Combination of the three ions were projected onto one another to emphasize similarities in their spatial distribution.

The lateral size was cropped to 5 μm× 5 μm to get rid of the errors due to the edge effect (atoms from the side walls can contribute to the measured signal). For an easier interpretation of the distribution of secondary ions in depth, the Z axis has been exaggerated 6 times, in reality this axis corresponds to a depth of 1.1 μm.

From the 3D models, we can observe the five consequent $Al_xGa_{1-x}In_yAs_{1-y}$– InP layers in the distribution of the three ions. This is in good agreement with the composition of the material (green dotted line in Figure 2). It is also clear that the bottom part (red dotted line in Figure 2) does not contain any $^{115}$In, and that there is a difference in thickness between the $^{27}$Al and $^{69}$Ga bottom layers due to the existing GaAs layer above the $Al_{0.9}Ga_{0.1}As$ layer. The intensities of $^{27}$Al and $^{69}$Ga ions are higher in this region compared to the upper layers.

In some regions secondary ions cannot be clearly distinguished (in example $^{115}$In inside the InP layers). This is due to the matrix effect, when interactions between the ion and its surroundings can substantially alter the yield and type of ions that we can observe.

## 5. Summary and conclusions

The biggest challenge for ToF-SIMS to be considered as an efficient 3D technique is to be able to accurately determine the depth from where the signal came from. The proposed combination of AFM and ToF-SIMS techniques takes us closer to achieve signal depth localization. It was demonstrated that consecutive sputtering and surface measuring can be successfully applied on materials which are ordered in an epitaxial manner. The concept of depth calibration of the SIMS depth profile by direct AFM depth measurements is demonstrated in this paper on a multilayer sample. SIMS-AFM scans of 10 μm × 10 μm area were alternated and controlled by a Python script.

Surface oxidation artefacts on SIMS profiles were corrected by averaging SIMS signals before and after AFM scans in order to reflect the real ion depth profiles. The concentration variations of $^{27}$Al, $^{69}$Ga and $^{115}$In ions in the layer structure of the VCSEL sample show good agreement with the expected composition, however without the depth calibration SIMS depth profiles do not coincide with the known structure. Sputter speed varies as a function of the local sample composition. We noticed higher sputter rates on the falling slopes of SIMS signals of the $^{115}$In-rich layers than on the corresponding rising slopes that indicates interlayer-mixing dependence of the sputter behaviour.

AFM scans showed bubble formation on the surface as a consequence of FIB milling. Ga droplet formation has been observed by measuring the root mean square roughness values. While both AFM scans and SEM imaging confirmed the droplets formation, SIMS data of Ga top projection showed no local increase in Ga ion intensity.

3-dimensional depth corrected models of the sputtered material were successfully reconstructed using SIMS layers corresponding to 20 nm thick slices. AFM data could also be used to correct small surface roughness changes, leading to the composition of real 3D data sets, but it would require that an AFM scan is taken after every ToF-SIMS frame, but this would imply signal variation of every ToF-SIMS scan due to surface oxidation.

It was out of the scope of this work to consider real 3D data set correction for small roughness changes (in the order of couple of tens of nanometers) in case of the VCSEL sample. For samples with a more complex surface topology, due to the sputter speed that depends on local composition variation having an impact on the SIMS signal, it would be mandatory to take into account the roughness effects for an adequate representation of the sample structure.

**Acknowledgement**

The present work was supported by ROTTOF project 5211.01371.900.01 and by the EM-PAPOSTDOCS-II programme (SzK), part of the European Union Horizon 2020 research and innovation programme under the Marie Slodowska-Curie grant agreement number 754364.